\documentclass[prb,twocolumn,amsmath,amssymb,showpacs]{revtex4-1} 

\pdfoutput=1
\usepackage{graphics}
\usepackage{dcolumn}
\usepackage{bm}
\usepackage{color}
\usepackage{epstopdf}
\usepackage{epsfig}
\usepackage[colorlinks,linkcolor=blue,citecolor=magenta]{hyperref}

\newcommand{\w}{\omega}
\newcommand{\s}{\sigma}
\newcommand{\bb}{\beta}
\newcommand{\al}{\alpha}

\newcommand{\De}{\Delta}
\newcommand{\de}{\delta}

\newcommand{\lam}{\lambda}
\newcommand{\ua}{\uparrow}
\newcommand{\dar}{\downarrow}

\newcommand{\f}[2]{\frac{#1}{#2}}
\newcommand{\expval}[1]{\left<#1\right>}
\newcommand{\nn}{\nonumber}
\newcommand{\imp}{\text{imp}}
\begin{document}
\title{Disorder and Zeeman coupling induced gap-filling states in the nodeless chiral superconducting Bi/Ni bilayer system }
	\author{Rasoul Ghadimi}
	\affiliation{Department of Physics, Sharif University of Technology, Tehran 14588-89694, Iran}
	\author{Mehdi Kargarian}
	\email{kargarian@physics.sharif.edu}
	\affiliation{Department of Physics, Sharif University of Technology, Tehran 14588-89694, Iran}
	\author{S. Akbar Jafari}
	\affiliation{Department of Physics, Sharif University of Technology, Tehran 14588-89694, Iran}
	\date{\today}
	\begin{abstract}
Motivated by the recently discovered time-reversal symmetry-breaking superconductivity in epitaxial Bi/Ni bilayer system with transition temperature $T_c\approx 4.2$K and the observation of zero-bias anomaly in tunneling measurements, we show that gap-filling states can appear in the fully gapped $d_{xy}\pm id_{x^2-y^2}$ superconducting states. We consider a model of helical electron states with d-wave pairing. In particular, we show that both magnetic and non-magnetic impurities can create states within the superconducting gap. Alternatively, we also show that the coupling of the electron spins to the in-plane Zeeman field provided by nickel can also create gap-filling states by producing Bogoliubov Fermi surfaces. Our findings may explain the origin of zero-bias anomaly observed in the point-contact tunneling measurements.         		
	\end{abstract}
	\maketitle
\section{Introduction}
	Interplay between topology and electronic band structures in insulators, superconductors, and metals has received enormous  attention and interest in recent years and has become a central issue in the condensed matter physics \cite{RevModPhys.82.3045,doi:10.1146/annurev-conmatphys-062910-140432,RevModPhys.83.1057}. The nontrivial topology encoded in the bulk band structures, usually characterized by integer numbers, yield nontrivial consequences for the electronic states living on the boundary of the system: the topologically protected gapless surface states. Helical edge states on the one-dimensional boundary of HgTe~\cite{Bernevig1757,Konig766}, Dirac electrons on the surface of three-dimensional topological materials Bi$_{2}$Se$_{3}$, Bi$_{2}$Te$_{3}$\cite{Zhang2009,Chen178}, Majorana fermions at the open ends of a one-dimensional topological superconductor\cite{Kitaev_2001}, and Femi arcs in topological Weyl semimetals\cite{PhysRevB.83.205101} such as TaAs\cite{PhysRevX.5.031013,Lv2015,Xu613} are a few known examples.           
	
Beside all, especial attentions have been paid to realization of topological superconductors promising a platform for topological quantum computations\cite{RevModPhys.80.1083}, a central paradigm in building a quantum computer. At the heart of material realization lies the nontrivial pairing wave functions of Cooper pairs of electrons in the vicinity of the Fermi surface. While the phonon-based mechanism for superconductivity leads to the conventional s-wave superconductor, the intrinsic spin or charge density fluctuations due to electron correlations may result in more complicated structure for pairing wave functions\cite{RevModPhys.78.17,RevModPhys.84.1383} for instance in cuprates, Sr$_2$RuO$_4$ \cite{0034-4885-79-5-054502,Luke1998},UPt$_3$\cite{Schemm190}. Alternatively the nontrivial pairing structures can be induced by an ingenious combination of more conventional materials. A famously celebrated structure has been introduced by Fu and Kane in Ref.[\onlinecite{PhysRevLett.100.096407}] where a conventional s-wave superconductor proximitized to the surface of topological insulators (sSc-TI interface) is shown to support Majorana zero-energy states in vortex cores. Yet, the surface of topological insulators can be replaced with more conventional two-dimensional electron gas with strong Rashba spin-orbit coupling, where the spin degeneracy is effectively lifted, and addition of a magnetic or Zeeman field turn the conventional induced s-wave superconductivity to a topological superconductor\cite{PhysRevLett.104.040502,PhysRevB.82.214509}.      

 \begin{figure}
 	\centering
 	\includegraphics[width=0.7\linewidth]{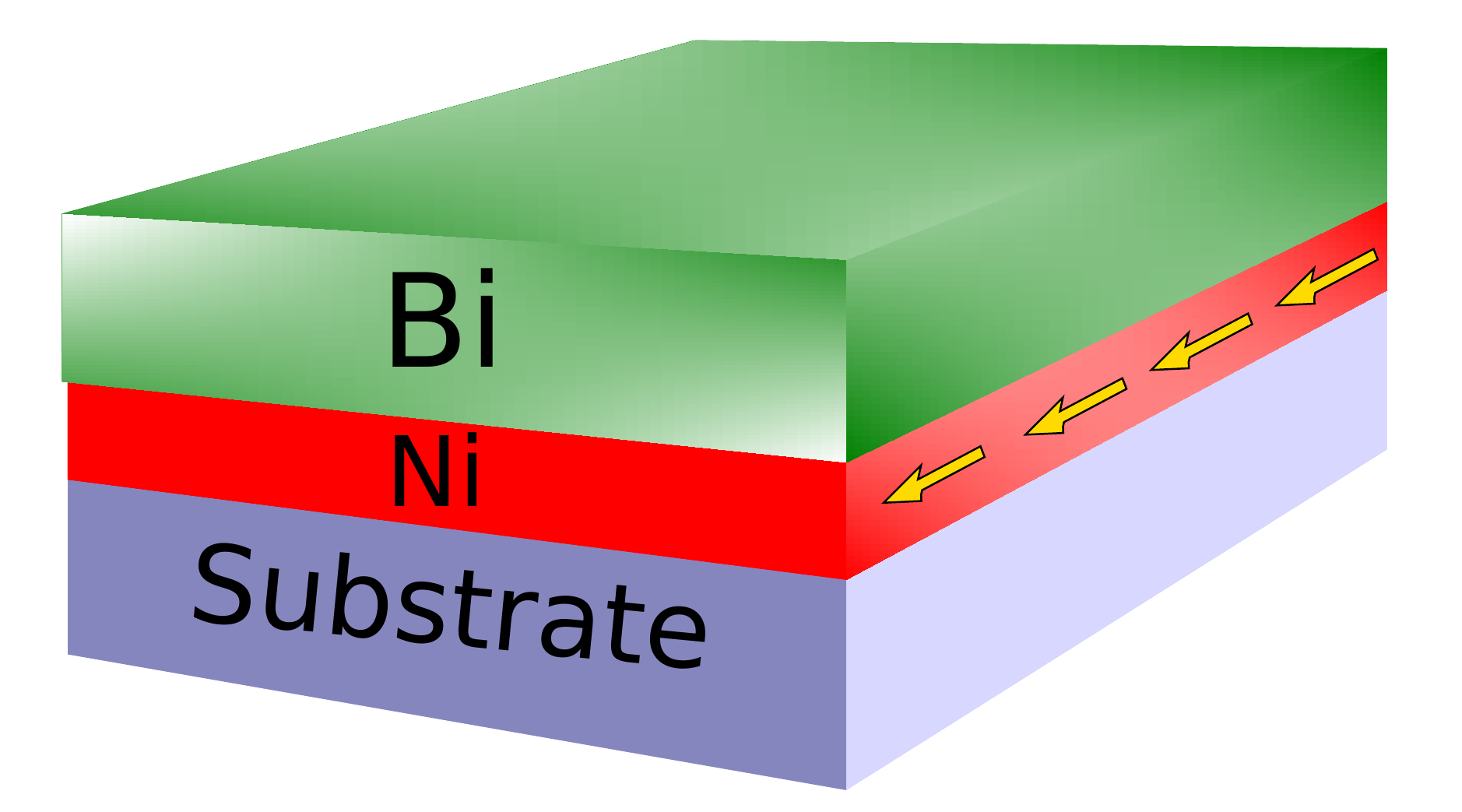}
 	\caption[BINI]{(Color online) Schematic representation of the epitaxial Bi/Ni bilayer system grown on a substrate such as MgO. The arrows show the in-plane magnetization of Ni.}
 	\label{fig:bini}
 \end{figure}
 
 The recently discovered superconductivity with rather high critical temperature $T_c\approx 4.2$K in epitaxial Bismuth-Nickel (Bi/Ni) bilayer system \cite{Gong2015} provided another example of a superconducting state with nontrivial pairing. The bilayer system is schematically shown in Fig.~\ref{fig:bini} where the surface termination Bi(110) is exposed to vacuum and Ni is a ferromagnet with in-plane magnetization. The optical measurements of Kerr signal from the Bi(110) side clearly shows that the superconducting states breaks the time-reversal symmetry (TRS) \cite{Gonge1602579}. Not only that, cooling below $T_c$ in a weak magnetic field applied to the sample in both directions and subsequently measuring resistivity on warming up and in zero field shows that the pairing states must break the TRS spontaneously. The theoretical model proposed in Ref.[\onlinecite{Gonge1602579}] uses the dominantly spin-orbit coupled electronic states localized on the surface of Bi(110) and shows that the pairing symmetry should be $d_{xy}\pm id_{x^2-y^2}$, a chiral topological superconductor characterized by Chern numbers $c=\pm2$. The complex structure of the superconducting order parameter indicates that the superconductivity in Bi/Ni system is fully gapped, an observation which is in agreement with the recently measured optical conductivity using time-domain THz spectroscopy \cite{PhysRevLett.122.017002}. The latter measurements further show that the superconductivity develops through the bulk of Bi with possible $p\pm ip$ pairing \cite{chao2018superconductivity}.             
 
 Beside the transport and optical measurements discussed above, a zero-bias anomaly has also been observed in the point-contact Andreev reflection \cite{Gong2015} which may seemingly contradict with having a fully gapped superconductor. This is exactly the motivation of our work in this paper. We ask the following specific question: how do the states filling the gap arise in a topologically nodless superconductor yielding a zero-bias anomaly? We give an affirmative answer to this question by introducing two possible scenarios. The gap-filling states appear as a result of (i) magnetic and non-magnetic impurities distributed randomly throughout, and/or (ii) emergence of a Bogolon Fermi surface due to the coupling of the electron spins to the in-plane magnetic moments as shown by arrows in Fig.~\ref{fig:bini}. In the second scenario we treat the disorder effects using the Abrikosov-Grokov formalism\cite{abrikosov1960aa} and ignore the inhomogeneity which might result in enhancement of superconductivity\cite{andersen2006thermodynamic,gastiasoro2013impurity,PhysRevB.98.184510,PhysRevLett.117.257002,PhysRevB.86.054507}. In particular, we show that there is a critical disorder strength beyond which the gap closes. The first scenario was first introduced for the sSC-TI interface and the gapless states emerge in the presence of a Zeeman field\cite{PhysRevB.97.115139}. We extend this finding to the chiral superconductors which may explain the observed zero-bias anomaly in Bi/Ni system. 
 
The paper is organized as follows. We first discuss the disorder-induced states in Sec~\ref{disorder} and then the effects of in-plane Zeeman coupling is discussed in Sec~\ref{LiFu}, and Sec.~\ref{onclusion} concludes. 

\section{disorder-induced gap-filling states in chiral superconductors}\label{disorder}
The electron states at the Bi(110) surface is strongly spin-momentum locked due breaking of inversion symmetry near the surface\cite{HOFMANN2006191} with a large hole pocket located around the center of surface Brillouin zone. Though our main motivation is to understand the origin of zero-bias anomaly in Bi/Ni system, in this section we consider two cases relevant to surface states of topological insulators and an 2D electron gas in the presence of Rashba spin-orbit coupling. We also assume that the magnetic fluctuations of Ni provide the pairing glue as detailed in Ref. [\onlinecite{Gonge1602579}]. We  study the effect of disorder on the superconducting states. 

 \subsection{Surface of Topological Insulator}
 The effective theory for the surface states of a topological insulators is proportional to the 2D Hamiltonian $\boldsymbol{\s}\times \mathbf{k}\cdot\hat{z}$, 
where $\mathbf{k}=(k_x,k_y)$ is the wave vector and $\boldsymbol{\s}=(\s_x,\s_y)$ are Pauli matrices\cite{RevModPhys.82.3045}.  
After a rotation it can be brought to the canonical Dirac form $\boldsymbol{\s}\cdot\mathbf{k}$, or equivalently,
 \begin{equation}
 H'_0=\sum_{\mathbf{k},\s\s'} c^\dag_{\mathbf{k}\s}\left[v_F |\mathbf{k}|(\sin\theta_{\mathbf{k}}\s_y+\cos\theta_{\mathbf{k}}\s_x)-\mu\right]_{\s\s'}c_{\mathbf{k}\s'},
 \label{helical.eqn}
 \end{equation}
where $\theta_{\mathbf{k}}$ is the polar angle of $\mathbf{k}$ plus $\pi/2$. 
 This Hamiltonian can be diagonalized by the following transformation
 \begin{equation}
 d^{\dag}_{\mathbf{k}\lam}=\frac{1}{\sqrt{2}}\left( c^{\dag}_{\mathbf{k}\ua}+\lam e^{i\theta_{\mathbf{k}}}c^{\dag}_{\mathbf{k}\dar}\right),
 \end{equation}
 where $\lam=\pm$ is helicity and label the energy bands.
Motivated by earlier work on the TR breaking superconductivity in Bi/Ni system and the theoretical proposal of $d\pm id$
in this system \cite{Gonge1602579}, let us formulate the effect of impurity scattering for such a superconducting state.
Using the time reversal (TR) operator $\Theta=i\s_y \mathcal{K}$, with $\mathcal{K}$ as complex conjugation operator, the TR transformation of creation operator $d^{\dag}_{\mathbf{k}\lam}$ becomes 
$\tilde d^{\dag}_{\mathbf{k}\lam}= \Theta d^{\dag}_{\mathbf{k}\lam} \Theta^{-1}=\lam e^{-i\theta_\mathbf{k}}d^{\dag}_{-\mathbf{k}\lam}$ which can be used to
construct the pairing interaction from time-reversed partners as  
 \begin{equation}
 H_{sc}= \sum_{\mathbf{k}} \De e^{-i2\theta_{\mathbf{k}}} d^{\dag}_{\mathbf{k}+}\tilde d^{\dag}_{\mathbf{k}+}+\mathrm{h.c.},
 \end{equation}
We have only included the pairing between the electron states with positive helicity which is justified
when the chemical potential is much larger than the pairing energy scale, i.e. the paring Hamiltonian is projected on the Fermi surface.   
In terms of the Nambu spinor $\psi_{\mathbf{k}}= ( d_{\mathbf{k}+}, \tilde d^{\dag}_{\mathbf{k}+})^T$ we can write the superconducting Hamiltonian as
\begin{equation}
H_0=\sum_{\mathbf{k}} \psi^{\dag}_{\mathbf{k}}\left[(v_F|\mathbf{k}|-\mu)\tau_3+\De\cos 2\theta_{\mathbf{k}}\tau_1+\De\sin 2\theta_{\mathbf{k}}\tau_2\right]\psi_{\mathbf{k}},
\end{equation}
where Pauli matrices $\tau$'s act within the particle-hole space. On top of this clean pairing Hamiltonian we add a disorder term $H_{\text{\imp}}$ such that the total Hamiltonian becomes,
\begin{equation}
H=H_0+H_{\text{\imp}}.
\end{equation}
The $H_{\imp}$ is given by
\begin{equation}
H_{\imp}=\sum_{i=1}^{N_{\text{\imp}}}\sum_{\s} c^{\dag}_{i\s}\left(\mu_i\de_{\s\s'}+J\mathbf{S}_i\cdot \boldsymbol{\s}_{\s\s'}\right)c_{i\s'}
\end{equation}
which describes scattering from magnetic and non-magnetic impurities. The random scalar potential is given by 
the random variables $\mu_i$, while the randomness in the magnetic impurities is determined by the random orientation
of the local spin $\mathbf{S}_i$. The exchange coupling $J$ is essentially determined by 
the hybridization of conduction and impurity electrons and the strength of the on-site Hubbard term which is assumed to 
be non-random. The average value of $\mu_i$ and $\mathbf{S}_i$ is zero, but their standard deviations are non-zero and together
with concentration of impurities determine the strength of interactions. 

By Fourier transform to momentum space the $H_{\imp}$ becomes 
 \begin{small}
 \begin{equation}
	H_{\imp}=\f{1}{V}\sum_{i,\s, \mathbf{k}\mathbf{k'}}e^{-i(\mathbf{k}-\mathbf{k'})\cdot\mathbf{R}_i}c^{\dag}_{\mathbf{k}\s}\left(\mu_i\de_{\s\s'}+J\mathbf{S}_i\cdot \boldsymbol{\s}_{\s\s'}\right)c_{\mathbf{k}'\s}\nn,
	\end{equation}
 \end{small}
and in band basis takes the following form
\begin{equation}
H_{\imp}=\f{1}{2V}\sum_{i,\lam\lam', \mathbf{k}\mathbf{k'}} d^{\dag}_{\mathbf{k}\lam} \bar{\mathcal{H}}_{\imp}(\mathbf{k},\mathbf{k}')d_{\mathbf{k}'\lam'},
\end{equation}
where 
\begin{equation}
\bar{\mathcal{H}}_{\imp}(\mathbf{k},\mathbf{k}')=e^{-i(\mathbf{k}-\mathbf{k'})\cdot\mathbf{R}_i}
	\left(\mu_iv_{\lam,\lam'}^{\mathbf{k},\mathbf{k'}}+ J\mathbf{S}_i\cdot \mathbf{m}_{\lam,\lam}^{\mathbf{k},\mathbf{k'}} \right)
\end{equation}
where $v_{\lam,\lam'}^{\mathbf{k},\mathbf{k'}}$ and $ \mathbf{m}^{\mathbf{k},\mathbf{k'}}$ are
\begin{gather}
v_{\lam,\lam'}^{\mathbf{k},\mathbf{k'}}\equiv
1+\lam\lam'e^{-i\theta_{\mathbf{k}}+i\theta_{\mathbf{k}'}}\nn\\
(\mathbf{m}_{\lam,\lam'}^{\mathbf{k},\mathbf{k'}})_x\equiv
\lam e^{-i\theta_{\mathbf{k}}}+\lam'e^{i\theta_{\mathbf{k}'}}\nn\\
(\mathbf{m}_{\lam,\lam'}^{\mathbf{k},\mathbf{k'}})_y\equiv
i\lam  e^{-i\theta_{\mathbf{k}}}-i\lam'e^{i\theta_{\mathbf{k}'}}\nn\\
(\mathbf{m}_{\lam,\lam'}^{\mathbf{k},\mathbf{k'}})_z\equiv
1-\lam\lam'e^{-i\theta_{\mathbf{k}}+i\theta_{\mathbf{k}'}}\nn
\end{gather}

The compact representation of the above Hamiltonian in the the Nambu space is,
\begin{equation} \label{Himp1}
H_{\imp}=\f{1}{2V}\sum_{i,\mathbf{k}\mathbf{k'}} \psi^{\dag}_{\mathbf{k}} \mathcal{H}_{\imp}(\mathbf{k},\mathbf{k}')\psi_{\mathbf{k}'},
\end{equation} where 
\begin{equation} \label{Himp2}
\mathcal{H}_{\imp}(\mathbf{k},\mathbf{k}')=e^{-i(\mathbf{k}-\mathbf{k'})\cdot\mathbf{R}_i}\left( \mu_iV^{\mathbf{k},\mathbf{k'}}+ J \mathbf{S}_i\cdot\mathbf{M}^{\mathbf{k},\mathbf{k'}}\right). 
\end{equation}

Here by assumption of large positive chemical potential we focus on $\lam=+1$ matrix elements, i.e.
\begin{eqnarray}
V^{\mathbf{k},\mathbf{k'}}=\left(\begin{matrix}
v_{++}^{\mathbf{k},\mathbf{k'}}&0\\0&\tilde v_{+,+}^{\mathbf{k},\mathbf{k'}}
\end{matrix}\right),\\
\mathbf{M}^{\mathbf{k},\mathbf{k'}}=\left(\begin{matrix}
\mathbf{m}_{++}^{\mathbf{k},\mathbf{k'}}&0\\0&\mathbf{\tilde m}_{++}^{\mathbf{k},\mathbf{k'}}
\end{matrix}\right),
\end{eqnarray}
where we have introduced,
\begin{eqnarray}
\tilde v_{\lam,\lam'}^{\mathbf{k},\mathbf{k'}}=(-\lam\lam'e^{-i\theta_{\mathbf{k}}+i\theta_{\mathbf{k}'}}) v_{\lam',\lam}^{-\mathbf{k}',-\mathbf{k}},\nn\\
\mathbf{\tilde m}_{\lam,\lam'}^{\mathbf{k},\mathbf{k'}}\equiv-\lam\lam'e^{-i\theta_{\mathbf{k}}+i\theta_{\mathbf{k}'}}\mathbf{m}_{\lam',\lam}^{-\mathbf{k}',-\mathbf{k}}.\nn
\end{eqnarray}

In order to study the effect of disorder, we use the language of Green's function. The disordered and clean Green function
are defined as $G(\mathbf{k},i\w_n)=(i\w_n- H)^{-1}$ and  $G^{(0)}(i\w_n)=(i\w_n-H_0)^{-1}$, respectively, where $i\w_n$'s are Fermionic Matsubara frequencies.
According to Dyson equation, these two are related by 
\begin{equation}\label{685545152152152}
G(\mathbf{k},i\w_n)^{-1}=  G^{(0)}(\mathbf{k},i\w_n)^{-1}-\Sigma(\mathbf{k},i\w_n),
\end{equation}
where in the self-consistent Born approximation up to second order, the self energy is given by\cite{PhysRevB.97.115139,bruus2004many}
\begin{align}
\Sigma(\mathbf{k},i\w_n)&\approx\expval{\mathcal{H}_{\imp}(\mathbf{k},\mathbf{k})}\nn \\
&+\sum_{\mathbf{k}'}\expval{\mathcal{H}_{\imp}(\mathbf{k},\mathbf{k}')G(\mathbf{k}',i\w_n)\mathcal{H}_{\imp}(\mathbf{k}',\mathbf{k})},
\label{scba.eqn}
\end{align}
where $\expval{\cdots}$ means the ensemble average is taken over disorder configurations. Following the approach presented in Refs. [\onlinecite{PhysRevB.94.214510}-\onlinecite{doi:10.1143/JPSJ.81.084707}] for treatment of disordered system, we assume that the excitations can be described in terms of an effective clean single-particle Hamiltonian $\tilde H_0$, which is similar to the original $H_{0}$ with $\Delta$ now depends on $\w_n$, i.e. we make the following replacement: $\Delta\to\Delta_n$ in $H_{0}$. Similarly for the Green function the spectral re-arrangements due to
disorder are taken into account by $i\w_n\to i\tilde\w_n$ where $i\tilde\w_n$ is a function of $i\w_n$ as described below. 
The effective Hamiltonian $\tilde H_0$ satisfies $G(i\w_n)=(i\tilde\w_n-\tilde H_0)^{-1}$. The Eq.~\eqref{685545152152152} and ~\eqref{scba.eqn} can now be cast into an algebraic self-consistency between $i\tilde\w_n$ and $\Delta_n$; See Eqs.~\eqref{omn.eqn}-\eqref{deln.eqn}. 

 Since we have assumed that the mean value of $\mu_i$ and $\mathbf{S}_i$ is zero, 
the first term in Eq.~\eqref{scba.eqn} gives zero. However their standard deviations are non-zero
and are given as follows: \cite{bruus2004many}
\begin{small}
\begin{gather}
\f{1}{V}\expval{u_iu_je^{-i\mathbf{q}\cdot R_i}e^{-i\mathbf{q'}\cdot R_j}}_{\imp}=n_{\imp}\bar{u^2}\de_{i,j}\de_{\mathbf{q},\mathbf{q'}},	\\
	\f{1}{V}\expval{S_i^{\al}S_j^{\bb}e^{-i\mathbf{q}\cdot R_i}e^{i\mathbf{q'}\cdot R_j}}_{\imp}= \f{1}{3}n_{\imp}S(S+1)\de_{ij}\de_{\al\bb}\de_{\mathbf{q}+\mathbf{q'},0},
\end{gather}
\end{small}

Obviously there is no correlation between electrostatic and magnetic scattering forces, $\langle \mu_i\mathbf{S}_j\rangle=0$. 
Therefore the non-zero contribution of the second term in Eq.~\eqref{scba.eqn} arises from the above fluctuations. By assumption of constant density of states (DOS) $N(0)$ at the Fermi surface the integral in the second term of Eq.~\eqref{scba.eqn} can be performed 
giving rise to the following equations for $i\tilde{\w}_n$ and $\De_n$:
\begin{eqnarray}
\tilde\w_n=\w_n+ \Gamma\f{\tilde\w_n}{2\sqrt{\tilde\De^2_{n}+\tilde{\w}_n^2}}\label{omn.eqn}\\
\tilde\De_n=\De +\alpha \Gamma\f{\tilde\Delta_n}{2\sqrt{\tilde\De^2_{n}+\tilde{\w}_n^2}} \label{deln.eqn}
\end{eqnarray}
where $\Gamma=\left(\bar{u^2}+S(S+1)J^2\right)\pi n_{\imp}N(0)$ defines the disorder strength. 
Note that the assumption of constant $N(0)$ around the Fermi level is justified for parabolic bands, as well as highly doped Dirac cone. 
For chiral $d\pm id$ superconductor, $\alpha=0$. This is because the $e^{2i\theta_{\mathbf{k}}}$ factors in the pairing enforce the angular integration
to become zero. Therefore in $d\pm id$ case, the pairing potential $\tilde\Delta_n$ does not change in the presence of disorder, in sharp contrast to s-wave pairing. For the case of s-wave pairing, and within a model that scatters single particles only (not the Cooper pairs), 
$\alpha=+1$\cite{doi:10.1143/JPSJ.81.084707} ($-1$\cite{PhysRevB.94.214510}) for non-magnetic (magnetic) disorders. 
This leads to an important difference between the chiral $d\pm id$ superconductivity and s-wave pairing. 
In the s-wave case, the fact that $\tilde\Delta_n$ depends on $n$, implies that for magnetic impurities ($\alpha=-1$) 
the $T_c$ will be suppressed much faster than the non-magnetic case ($\alpha=+1$). 
In the chiral $d\pm id$ case, the $\alpha$ is zero anyhow, and to that extent, magnetic and non-magnetic impurities
have comparable effect on the $T_c$. As a function of impurity concentration, the transition temperature reads as\cite{PhysRevB.94.214510, doi:10.1143/JPSJ.81.084707} 
\begin{equation} T_{c}(n_{i})=T_{c}(0)-\frac{\pi}{8}(1-\alpha)\Gamma. \end{equation}
Therefore the chiral superconductivity is fragile against 
impurity scattering, irrespective of the magnetic nature of scattering center. 

\begin{figure}[t]
	\centering
	\includegraphics[width=0.9\linewidth]{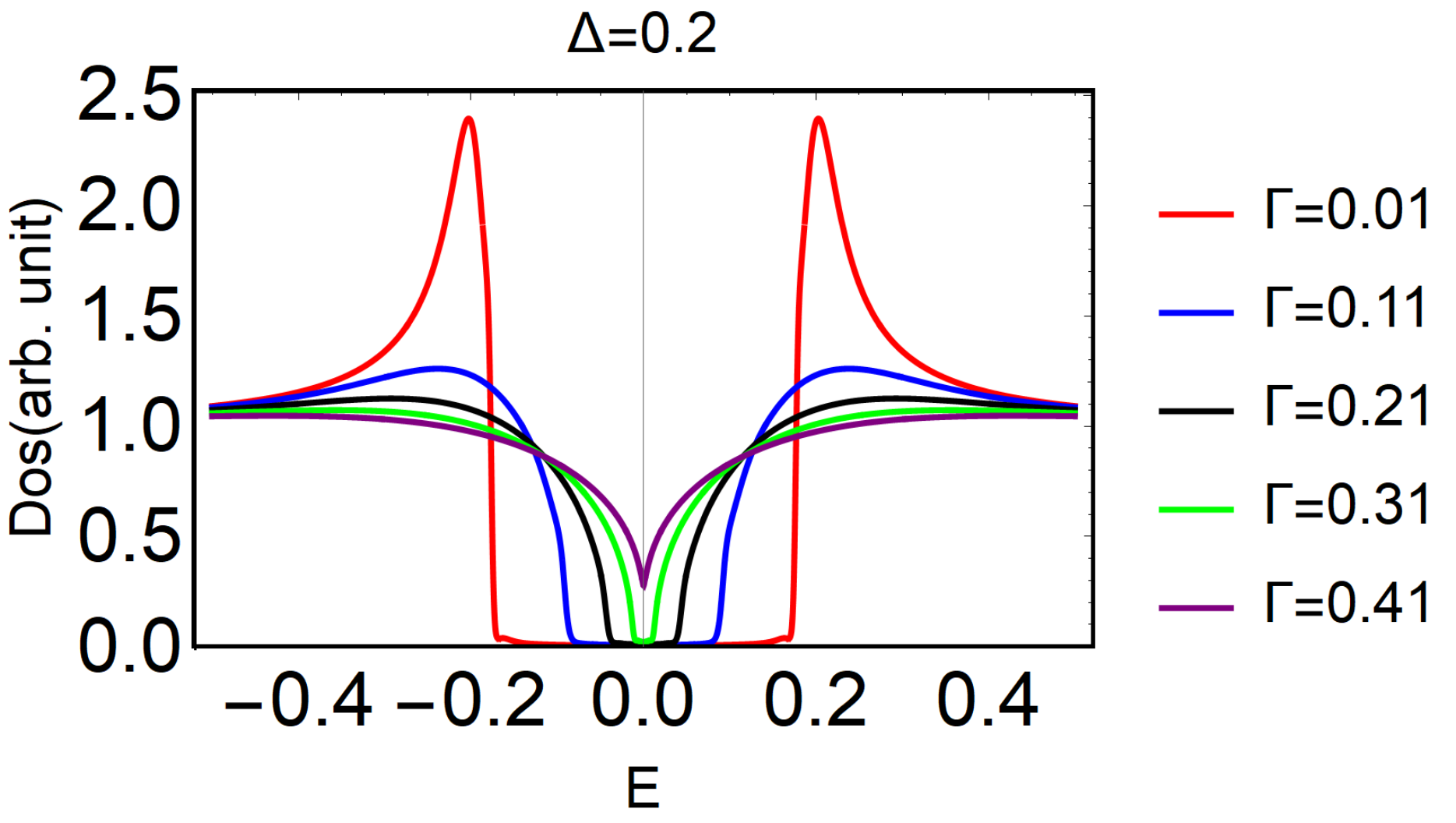}
	\caption{Density of states as a function of energy for different disorder strength $\Gamma$ by $d+id$ superconductivity with amplitude $\De=0.2$, show spectral gap closing with increasing the disorder strength}
	\label{fig:disordertoplogicalsurfacedpid}
\end{figure}

The DOS can be found by momentum integration after analytical continuation $i\w_n\to \omega+i0^+$ 
of the Green function yielding \begin{equation}
N(\w)=-\f{1}{\pi}N(0)\mathrm{Im}\left[\lim_{i\w_n\to\w+i0^+}\f{-i\tilde{\w}_n}{\sqrt{\tilde \w_n^2+ \De^2}}\right].
\end{equation}
One way to perform the analytical continuation is to use the Pad\'e approximation to represent $-i\tilde{\w}_n/\sqrt{\tilde \w_n^2+ \De^2}$ by a function 
$F(i\w_n)=Q(i\w_n)/P(i\w_n)$, where $Q$ and $P$ are polynomials of $i\w_n$, from which its analytical continuation $F(\w+i0^+)$ can be evaluated. Alternative way is to perform the analytical continuation before solving 
the self-consistency equations~\eqref{omn.eqn} and~\eqref{deln.eqn} \cite{fathi2010dynamical}. We have checked that
these two approaches give identical results. 

\begin{figure}
	\centering
	\includegraphics[width=0.49\linewidth]{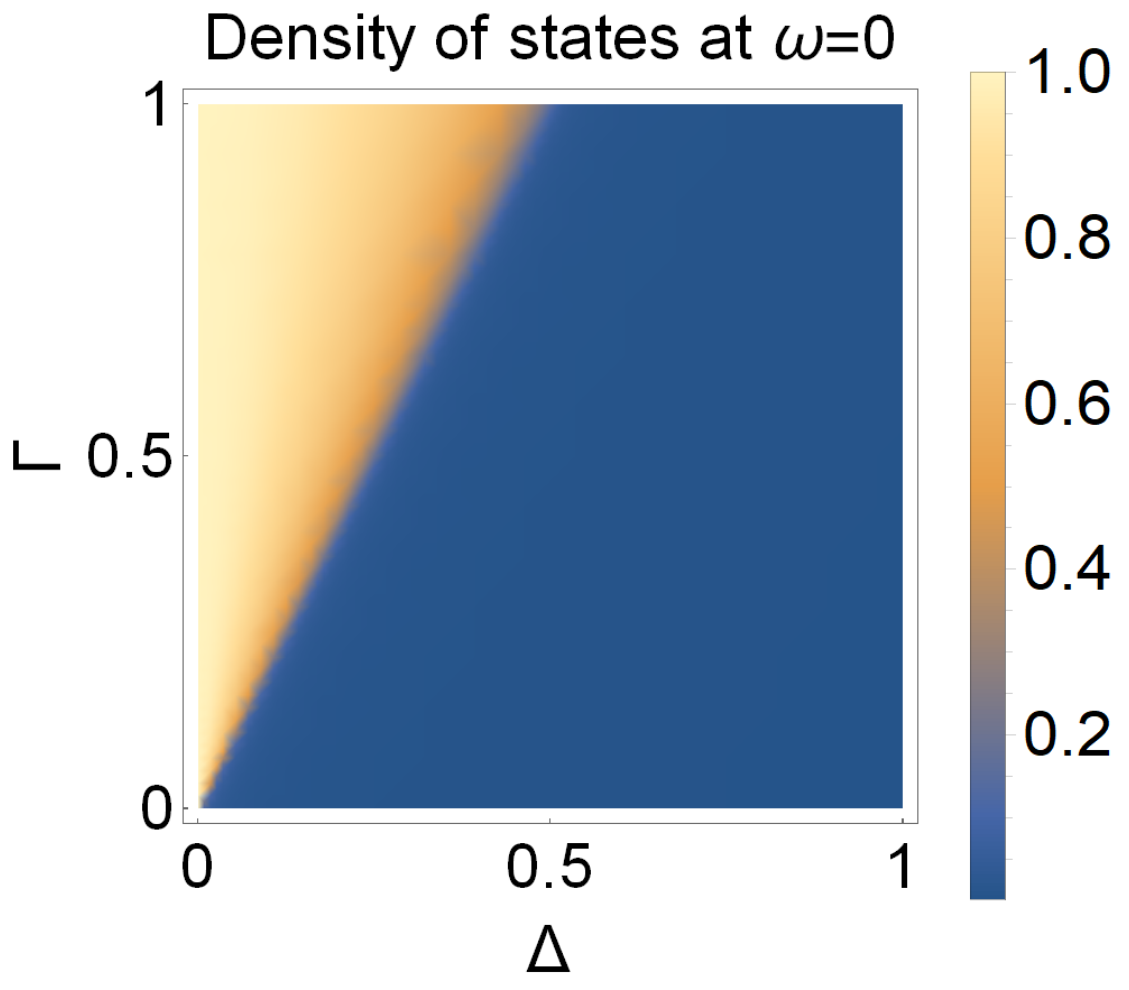}
	\includegraphics[width=0.49\linewidth]{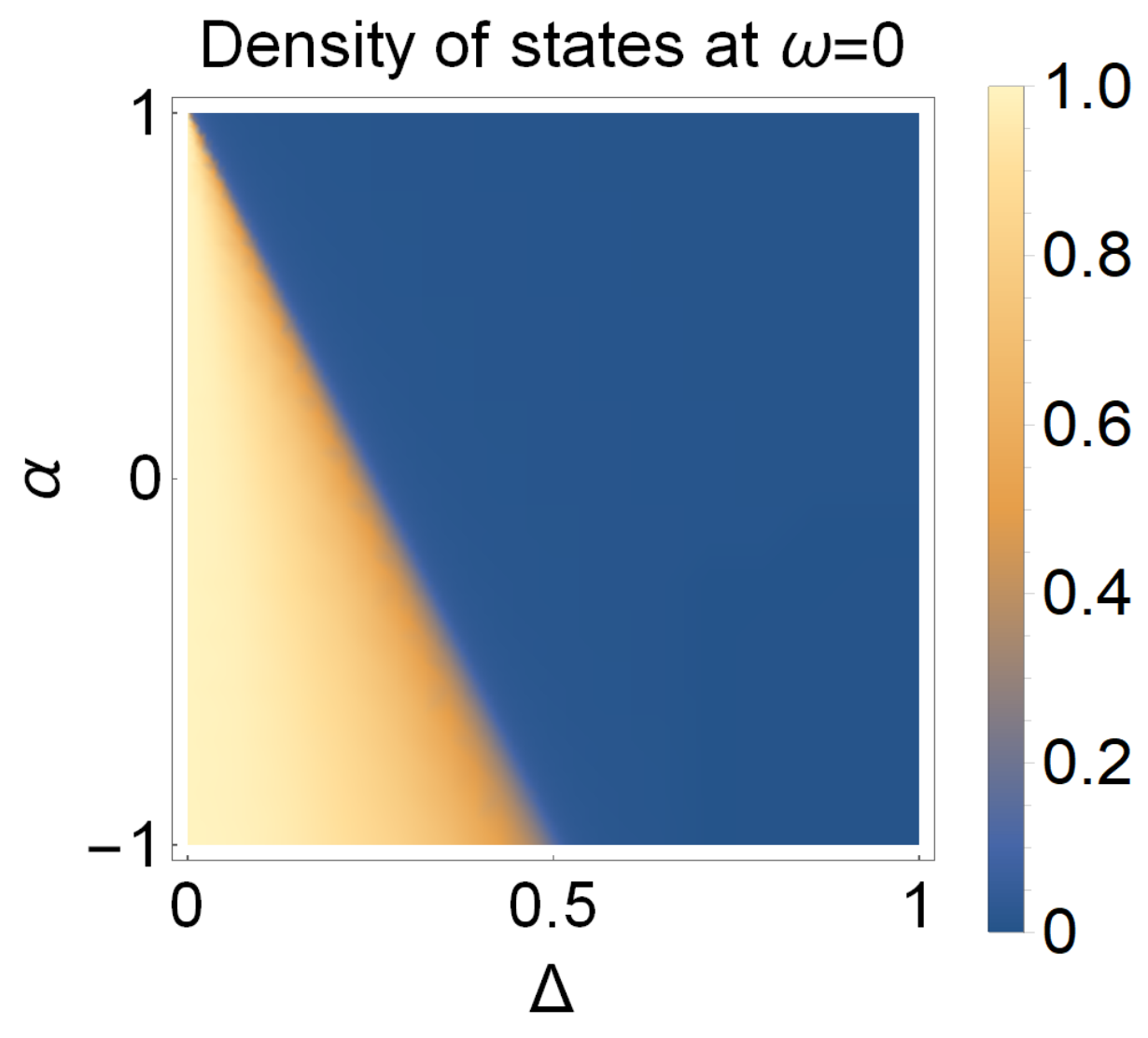}
	\caption{(left) Density plot of  $\w=0$  density of states in terms of disorder and pairing potential strength.
	(right) Map of the DOS at $\omega=0$ for relatively large disorder strength $\Gamma=0.5$ in the plane of 
	$\alpha$ and $\Delta$. $\alpha=0$ corresponds to chiral superconductivity while $\alpha=+1$ ($\alpha=-1$) corresponds to
	$s$-wave superconductor with non-magnetic (magnetic) scatterers.}
	\label{comparison.fig}
\end{figure}
In Fig.~\ref{fig:disordertoplogicalsurfacedpid}, we have plotted the DOS for $\De=0.2$ as a function of energy  
for different disorder strength $\Gamma$ indicated in the legend. As can be seen, beyond a certain critical disorder strength 
the superconducting gap is completely filled and we will have a gapless superconductor. 
The value of DOS at $\omega=0$ determines whether the superconductor is gapless or gapped. 
Therefore in left panel of Fig.~\ref{comparison.fig} we probe the density of states for $\w\to0$
in the plane of pairing and disorder strength. This gives us a border that separate gapless and gapped superconducting phases.
To emphasize the difference between the behavior of chiral and non-chiral superconductors against disorder, in right panel
of Fig.~\ref{comparison.fig} for a fixed disorder strength of $\Gamma=0.5$, we have generated a map of the DOS at $\omega=0$ in the
plane of $\alpha$ and $\Delta$. The chiral superconductor corresponds to $\alpha=0$. The s-wave superconductors with magnetic (non-magnetic)
scatterers correspond to $\alpha=-1$ ($\alpha=+1$). This comparison indicates that the chiral superconductor is capable of producing
gapless superconductivity much easier than s-wave superconductor. 
This can be a possible explanation for the observed in-gap features in the Bi/Ni system \cite{Gong2015}.

\subsection{Electron Gas with Rashba Coupling}
In this section we use the methods used in preceding subsection to study the affect of disorder on chiral superconductors 
in system with multiple Fermi surfaces. For simplicity we consider a two-dimensional electron gas in the presence of a strong Rashba
spin-orbit coupling described by the following Hamiltonian     
 
\begin{equation}
H_0=\sum_{\mathbf{k},\s\s'} c^\dag_{\mathbf{k}\s}H_{\s\s'}(\mathbf{k}) c_{\mathbf{k}\s'}
\label{rashba.eqn}
\end{equation}
with 
\begin{equation} 
H_{\s\s'}(\mathbf{k})=\left[\f{|\mathbf{k}|^2}{2m}\de_{\s\s'}+v_R |\mathbf{k}|(\sin\theta_{\mathbf{k}}\s_y+\cos\theta_{\mathbf{k}}\s_x)-\mu\right]_{\s\s'},
\end{equation} 
where the spin-orbit coupling is expressed in the form of a velocity scale $v_R$ to make it look similar
to the helical surface states considered in the preceding section. This Hamiltonian reduces to the 
helical Hamiltonian~\eqref{helical.eqn} upon replacement $m\to\infty$ and $v_R\to v_F$. 
The pairing in the two cases is however
different. In the helical metallic case for any chemical potential, there is only one Fermi contour. 
But in the present Rashba spin-orbit coupled case, the spin-orbit coupling splits them into
two Fermi contours. For large enough chemical potential, the two spin-orbit split Fermi contours
will have opposite helicity. Therefore the pairing will include {\it both} helicities $\lambda=\pm 1$
\begin{equation}
H_{sc}= \sum_{\mathbf{k},\lambda} \De_{\lam} e^{-i2\theta_{\mathbf{k}}} d^{\dag}_{\mathbf{k}\lam}\tilde d^{\dag}_{\mathbf{k}\lam}+\mathrm{h.c.}, 
\end{equation}
which together with Eq.~\eqref{rashba.eqn} yields the following superconducting Hamiltonian 
\begin{widetext}
		\begin{eqnarray}
		H=\sum_{\mathbf{k}\lam} \psi^{\dag}_{\mathbf{k}}\left(\f{\s_0+\lam \s_3}{2}\right)\otimes\left[\left(\f{|\mathbf{k}|^2}{2m}+\lam v_R|\mathbf{k}|-\mu\right)\tau_3+\De_{\lam}(\cos 2\theta_{\mathbf{k}}\tau_1+\sin 2\theta_{\mathbf{k}}\tau_2)\right]\psi_{\mathbf{k}}.
		\end{eqnarray}
\end{widetext}
where $\psi_{\mathbf{k}}= (
d_{\mathbf{k}+},\tilde d^{\dag}_{\mathbf{k}+},
d_{\mathbf{k}-},\tilde d^{\dag}_{\mathbf{k}-}
)^T$ is the Nambu spinor.

The calculation proceeds along the same steps as preceding subsection, with a difference that
instead of one helicity ($\lambda=+$ for the surface of TI), now will have both helicities $\lambda=\pm$
and hence for the impurity Hamiltonians in Eqs.~\eqref{Himp1}-\eqref{Himp2} the relevant matrices will be $4\times 4$ which are given by
\begin{equation}
V^{\mathbf{k},\mathbf{k'}}=\left(\begin{matrix}
v_{++}^{\mathbf{k},\mathbf{k'}}&0&v_{+-}^{\mathbf{k},\mathbf{k'}}&0
\\0&\tilde v_{+,+}^{\mathbf{k},\mathbf{k'}}&0&\tilde v_{+,-}^{\mathbf{k},\mathbf{k'}}
\\v_{-+}^{\mathbf{k},\mathbf{k'}}&0&v_{--}^{\mathbf{k},\mathbf{k'}}&0
\\0&\tilde v_{-+}^{\mathbf{k},\mathbf{k'}}&0&\tilde v_{--}^{\mathbf{k},\mathbf{k'}}
\end{matrix}\right),
\end{equation}
and
\begin{equation}
\mathbf{M}^{\mathbf{k},\mathbf{k'}}=\left(\begin{matrix}
\mathbf{m}_{++}^{\mathbf{k},\mathbf{k'}}&0&\mathbf{m}_{+-}^{\mathbf{k},\mathbf{k'}}&0
\\0&\mathbf{\tilde m}_{++}^{\mathbf{k},\mathbf{k'}}&0&\mathbf{\tilde m}_{+-}^{\mathbf{k},\mathbf{k'}}
\\\mathbf{m}_{-+}^{\mathbf{k},\mathbf{k'}}&0&\mathbf{m}_{--}^{\mathbf{k},\mathbf{k'}}&0
\\0&\mathbf{\tilde m}_{-+}^{\mathbf{k},\mathbf{k'}}&0&\mathbf{\tilde m}_{--}^{\mathbf{k},\mathbf{k'}}
\end{matrix}\right).
\end{equation}
Again due to a vanishing angular integration coming from the $e^{2i\theta_{\mathbf{k}}}$, for chiral $d\pm id$ 
superconductor we obtain $\tilde\De_{\lam,n}= \De_{\lam}$ which again corresponds to $\alpha=0$ situation 
in Eq.~\eqref{deln.eqn}, albeit with new helicity index acquired due to two Fermi contours. 
The self consistency equation corresponding to Eq.~\eqref{omn.eqn} will now become
\begin{equation}
\tilde \w_n=\w_n+\Gamma\left( \xi\f{\tilde{\w_n}}{2\sqrt{\tilde \w_n^2+ \De_{-}^2}}+ \f{\tilde{\w_n}}{2\sqrt{\tilde \w_n^2+ \De_{+}^2}}\right),
\label{omn2.eqn}
\end{equation}
where 
$$N_{\lam}(0)=\frac{m}{2\pi}\left(1-\lam\frac{v_{R}}{\sqrt{\frac{2\mu}{m}+v_{R}^2}} \right) $$ is the DOS for Fermi contour with helicity $\lam$,   
$$\xi=\f{N_{-}(0)}{N_{+}(0)}>1.$$
 
\begin{figure}
	\centering
	\includegraphics[width=1\linewidth]{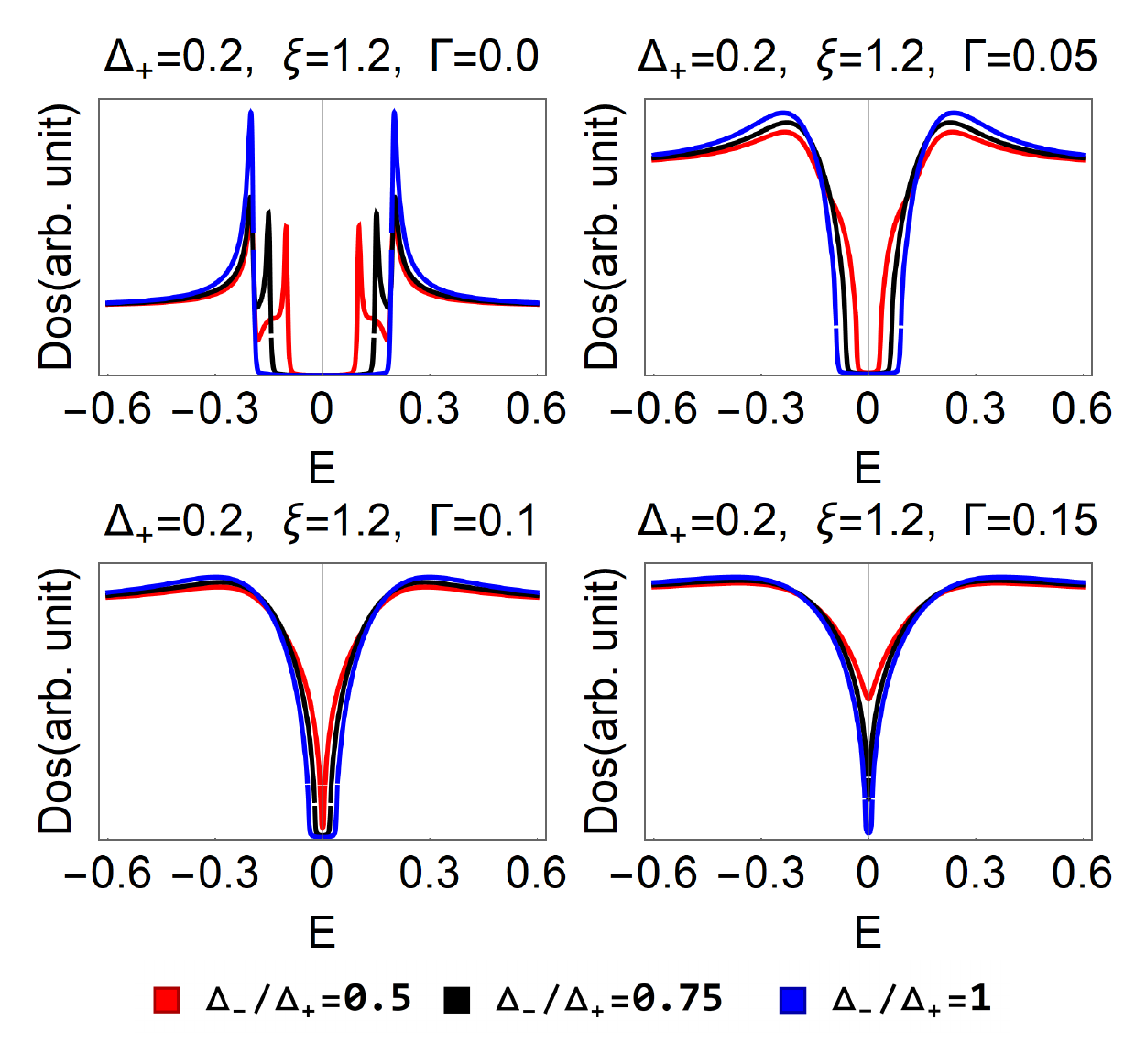}
	\caption{Density of states for  different $\De_{-}/\De_{+}$ ratio and $\xi=1.2$ as a function of energy plotted for (a)$\Gamma=0$(b)$\Gamma=0.05$(c)$\Gamma=0.1$(d)$\Gamma=0.15$}
	\label{fig:withoutdisorderbychangingdeltplusandminus}
\end{figure}

In this case from Eq.~\eqref{omn2.eqn} it follows that the total DOS is given by 
\begin{equation}
N(\w)=\f{2}{\Gamma\pi}\mathrm{Im}\left[\lim_{i\w_n\to\w+i0^+}i\left(\tilde \w_n-\w_n\right)\right],
\end{equation}
which again can be calculated as before, either by Pad\'e analytic continuation, or
direct solution of the self-consistency equations slightly above the real frequency axis\cite{fathi2010dynamical}. Density of states for different $\De_{+}/\De_{-}$ ratio with $\De_+=0.2$ and $\xi=1.2$ as a function of energy plotted in Fig.~\ref{fig:withoutdisorderbychangingdeltplusandminus}  for different disorder strength (a) $\Gamma=0$, (b) $\Gamma=0.05$, (c) $\Gamma=0.1$ and (d) $\Gamma=0.15$. 
In panel (a) where the disorder is zero, both gaps are clearly visible as two coherence peaks. 
Adding a little bit of disorder, in panel (b), the visible peaks are smeared out and both gaps start to get filled. Indeed within Eq.~\eqref{omn2.eqn}, if the DOS at zero for the two bands would be the same ($\xi=1$) then both gaps would evolve similarly. In the present case corresponding to $\xi=1.2$, both gaps are filled almost in the same
manner. In this way, the smaller of the two gaps will be filled in first. This is what happens in panel (c) where the smaller gap is filled in, while the
larger gap still survives. Finally in panel (d), both gaps are filled. Transport measurements can clearly indicate the value of $\Gamma$ at which the
smaller gap is filled. This is a disorder strength at which zero-bias conductance takes off from zero. The second critical disorder at which 
the larger gap is filled, shows up as a kink in the trend of zero-bias peak as a function of disorder. The presence of such a kink is an
essential difference between the chiral superconductivity in helical states and Rashba spin-orbit coupled systems. If the system is in a regime
where the smaller gap is already filled, specific heat measurement will detect the in-gap features {\it and} the larger gap. 

\begin{figure*}
	\centering
	\includegraphics[width=0.8\linewidth]{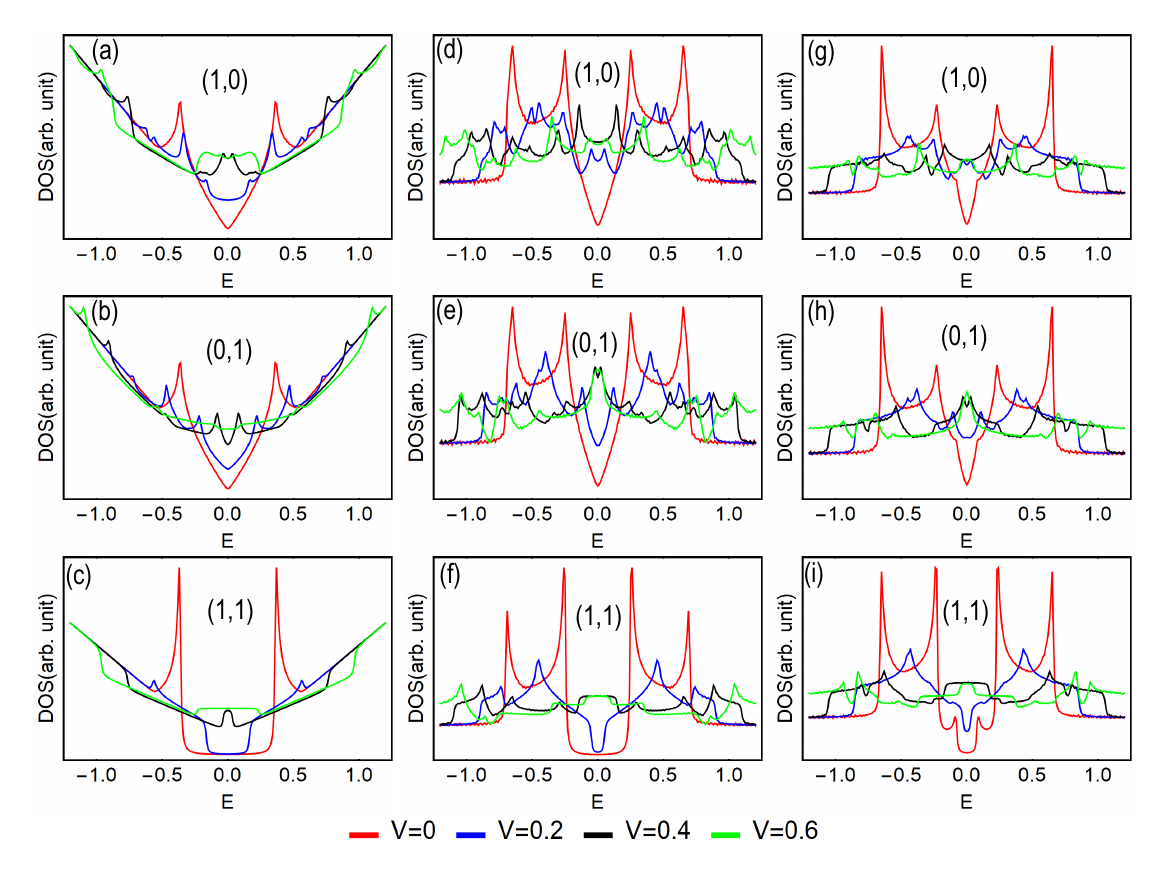}
	\caption{Density of states for the the surface of topological insulator (first column, a,b,c) and Rashba electron gas with 
	spin-singlet $(\De_0,\De_1)=(0.5,0)$ (second column, d,e,f) and spin-triplet $(\De_0,\De_1)=(0,0.5)$ (third column, g,h,i) 
	for the $d$-wave superconductivity.
	The contribution of $d_{x^2-y^2}$ and $d_{xy}$ pairing in each row is as follows:
	In first row (a,e,h) $(\zeta_{x^2-y^2},\zeta_{xy})=(1,0)$. In second row (b,f,i) $(\zeta_{x^2-y^2},\zeta_{xy})=(0,1)$
	and in third row (c,g,j) we have set $(\zeta_{x^2-y^2},\zeta_{xy})=(1,1)$ shown in each panel. The values of
	in-plane Zeeman field $V$ is indicated in the legend.	}\label{fig:LiangFu}
\end{figure*}

\section{Gap-Filing states caused by in-plain Zeeman coupling}\label{LiFu}
In this section we investigate the effect of in-plane magnetic field as another possible origin of mid-gap states. 
This mechanism was suggested recently~\cite{PhysRevB.97.115139} in the context of
sSc-TI. In this section we investigate this mechanism
for the chiral $d\pm id$ superconductivity and also for the $d$-wave pairings. In our set-up shown in Fig.~\ref{fig:bini} the in-plane Zeeman coupling can be provided by the proximity to Ni. We start by adding an in-plane Zeeman field $V\sigma_y$ to Hamiltonian~\eqref{rashba.eqn} 
\begin{small}
	\begin{gather}\label{TILIANGFU7}
		H_0=\sum_{\mathbf{k}} c^\dag_{\mathbf{k}}\left[\f{|\mathbf{k}|^2}{2m}\s_0+v_R(k_x\s_y-k_y\s_x)-\mu-V\s_y\right]c_{\mathbf{k}}\nn\\
		+\sum_{\mathbf{k},\s\s'}\left(\De_{\s\s'}(\mathbf{k})c^\dag_{\mathbf{k},\s}c^\dag_{-\mathbf{k},\s'}+\text{h.c.}\right),
	\end{gather}
\end{small}
where the matrix $\De(\mathbf{k})$ is parameterized as follows,
\begin{equation}\label{pairingmatrix}
\De(\mathbf{k})=
\left(\begin{matrix}
i|\mathbf{k}|\De_1(\mathbf{k}) e^{-i\theta_{\mathbf{k}}}&\De_0(\mathbf{k})\\-\De_0(\mathbf{k}) &i|\mathbf{k}|\De_1(\mathbf{k}) e^{i\theta_{\mathbf{k}}}
\end{matrix}\right).
\end{equation}
In following for numerical calculations we set $\left(m,v_R,\mu\right)=\left(0.3,1,0.5\right)$. 
To address the surface of topological insulator we let $1/2m\to 0$ and $v_R=v_F$. 

For the Rashba model with two Fermi contours $\lam=\pm$, and consequently pairings $\De_{\pm}(\mathbf{k})$, the gap functions in Eq.~\eqref{pairingmatrix} can be written as\cite{samokhin2008gap,ghadimi2018competing} $\De_{1}(\mathbf{k})=\De_{+}(\mathbf{k})-\De_{-}(\mathbf{k})$ and $\De_{0}(\mathbf{k})=\De_{+}(\mathbf{k})+\De_{-}(\mathbf{k})$. We take a general case for band pairings as $\De_{\pm}(\mathbf{k})=(\De_{0}\pm \De_{1})\left(\zeta_{x^2-y^2} \cos 2\theta_{\mathbf{k}}+i \zeta_{xy} \sin 2\theta_{\mathbf{k}}\right)$, where $(\zeta_{x^2-y^2},\zeta_{xy})$ are ${\cal O}(1)$ numerical
parameters that encode the relative contribution of $d_{x^2-y^2}$ and $d_{xy}$ pairings. With this identification for pairing functions, we rewrite the pairing matrix (\ref{pairingmatrix}) as
\begin{equation}\label{pairingmatrix2}
\left(\begin{matrix}
i|\mathbf{k}|\De_1 e^{-i\theta_{\mathbf{k}}}&\De_0\\-\De_0 &i|\mathbf{k}|\De_1 e^{i\theta_{\mathbf{k}}} 
\end{matrix}\right)\left(\zeta_{x^2-y^2} \cos 2\theta_{\mathbf{k}}+i \zeta_{xy} \sin 2\theta_{\mathbf{k}}\right)
\end{equation}
where here $(k_x,k_y)=|\mathbf{k}|(\cos\theta_{\mathbf{k}},\sin\theta_{\mathbf{k}})$. In this way the explicit $d\pm id$ structure
has been factored out. Now the amplitudes $\Delta_0$ and $\Delta_1$ do not depend on $\mathbf{k}$. 
The above pairing matrix has the structure that the superconductivity is nodal unless both $\zeta_{x^2-y^2}$ and $\zeta_{xy}$ become nonzero, 
which then breaks the TRS. 

In Fig.~\ref{fig:LiangFu} we present DOS for various cases. The three rows, from top to bottom, correspond to $(\zeta_{x^2-y^2},\zeta_{xy})=(1,0)$, 
$(\zeta_{x^2-y^2},\zeta_{xy})=(0,1)$ and $(\zeta_{x^2-y^2},\zeta_{xy})=(1,1)$, respectively. The panels in the first column depict the results for the surface of topological insulators, while those in next two columns are for the electron gas with Rashba coupling.      

Let us first consider the surface of topological insulator. As pointed out, in this case we need to set $\De_0=\De_1=0.5$. The corresponding DOS is shown in panels (a), (b), and (c) in Fig.~\ref{fig:LiangFu}. The first two panels, (a) and (b), correspond to nodal $d_{x^2-y^2}$ and  $d_{xy}$ pairing symmetries. As seen, in the absence of the Zeeman field we have a V-shaped pseudo-gap arising from the node in the pairing function as shown by solid red line. A smallest amount of in-plane Zeeman $V$ gives rise to Bogoliubov Fermi contours, and consequently a finite DOS is created at $E=0$. The most interesting situation occurs when the superconducting state is fully gapped (panel (c)) due to TRS breaking of the $d+id$ pairing wave function. In the absence of the Zeeman coupling there is clear gap in the DOS. However, a rather strong Zeeman coupling can fill up the gap by creating Bogoliubov Fermi contours.     

For the electron gas with Rashba coupling we can also tune $\De_{0}$ and $\De_{1}$ independently. We consider two extreme cases where $(\De_{0},\De_{1})=(1,0)$ correspond to purely singlet pairing as shown in panels (d), (e), and (f), and $(\De_{0},\De_{1})=(0,1)$ represent the triplet case in panels (g), (h), and (i). Again it is clearly seen that for nodal cases a tiny Zeeman field creates finite DOS at zero energy, and the fully gapped case needs stronger Zeeman coupling for the gap to be filled with Bogoliubov quasiparticles.

\section{Conclusion\label{conclusion}}
This work is mainly motivated by the observation of zero-bias anomaly in the point-contact tunneling measurements in the nodless time-reversal symmetry-breaking superconducting epitaxial Bi/Ni bilayer system. In the previous work by one of the authors and coworkers in Ref.[\onlinecite{Gonge1602579}], it was proposed that the pairing symmetry should be the surface chiral $d\pm id$, in agreement with Kerr measurements and thickness-dependent of transition temperature. In this work we considered the effects of both magnetic and non-magnetic impurities on the superconducting gap. We showed in all cases mid-gap states appear in the gap and start to fill it up. We also showed that the in-plane Zeeman coupling, which is produced by Ni lying underneath Bi, can also create Bogoliubov Fermi contours of quasiparticles with non-vanishing spectral weight within the gap. We believe our finding may explain the anomalous zero-bias signal observed.

\section{Acknowledgements}
M. K. acknowledges the support from the Sharif University
of Technology under Grant No. G690208.
S. A. J. acknowledges grant number G960214 of the research deputy, Sharif University of Technology.
%
	
\end{document}